\documentclass[twocolumn,aps,prl,showpacs, superscriptaddress, amsmath,amssymb]{revtex4}

\usepackage{graphicx}
\usepackage{dcolumn}
\usepackage{bm}
\usepackage{epstopdf}

\begin{document}

\title{Impact dynamics of granular jets with non-circular cross-sections}
\author{Xiang Cheng}
\affiliation{Department of Chemical Engineering and Materials
Science, University of Minnesota, Minneapolis, MN 55455}
\affiliation{The James Franck Institute and Department of Physics,
The University of Chicago, Chicago, Illinois 60637}
\author{Leonardo Gordillo}
\affiliation{The James Franck Institute and Department of Physics,
The University of Chicago, Chicago, Illinois 60637}
\affiliation{Laboratoire ``Mati{\`e}re et Syst{\`e}mes Complexes''
(MSC), UMR 7057 CNRS, Universit{\'e} Paris 7 Diderot, 75205 Paris
Cedex 13, France}
\author{Wendy W. Zhang}
\affiliation{The James Franck Institute and Department of Physics,
The University of Chicago, Chicago, Illinois 60637}
\author{Heinrich M. Jaeger}
\affiliation{The James Franck Institute and Department of Physics,
The University of Chicago, Chicago, Illinois 60637}
\author{Sidney R. Nagel}
\affiliation{The James Franck Institute and Department of Physics,
The University of Chicago, Chicago, Illinois 60637}

\date{\today}

\begin{abstract}

Using high-speed photography, we investigate two distinct regimes of the impact
dynamics of granular jets with non-circular cross-sections.  In the steady-state regime, we observe the
formation of thin granular sheets with anisotropic shapes and show
that the degree of anisotropy increases with the aspect ratio of the
jet's cross-section. Our results illustrate the
liquid-like behavior of granular materials during impact and
demonstrate that a
collective hydrodynamic flow emerges from strongly interacting
discrete particles. We discuss the analogy between our experiments and those from the Relativistic Heavy Ion Collider (RHIC), where similar anisotropic ejecta from a quark-gluon plasma have been observed in heavy-ion impact.  

\end{abstract}

\maketitle

\section{I. Introduction}

When a liquid jet impacts on a solid surface, it shatters violently,
spreads out rapidly and deforms into an elegant thin film \cite{1}.
Such an impressive phenomenon has attracted both lay people and
scientists. In 1833, well before the invention of
high-speed photography, the French physicist Savart conducted a ``water-bell'' experiment and illustrated with precise hand-drawn pictures the dynamics of a
water jet hitting a small disc \cite{2}. The ubiquitous fluid impact process, occurring in many
natural and industrial circumstances, is determined by
momentum and energy transfer between the fluid and
the solid surface, and depends on fluid properties such as inertia,
surface tension and viscosity \cite{3,4}.

The pattern of ejecta created by impact has been used to reveal some fluid-like properties in different materials.  For example,
it has been found that a jet of granular materials hitting a solid
target can form a hollow cone or a thin sheet, with a shape
quantitatively matching that of a liquid {\it without} surface
tension \cite{5,6,7}.  As another more esoteric example, the
coherent ejecta patterns observed at the Relativistic Heavy Ion
Collider (RHIC) have been analyzed to indicate that the quark-gluon
plasma generated from impact of relativistic ions is a nearly perfect
(zero-shear-viscosity) liquid \cite{8,9}. In spite of 16 orders of
magnitude difference in their energy scales, the two examples share
one common feature: they both illustrate emergent hydrodynamic flows
out of strongly interacting discrete particles. Thus, granular jet
impact has been studied in part as a macroscopic analog of the RHIC
experiment \cite{5,10,11} or as a fluid with a finite viscosity \cite{12,13}.  Indeed,
similar to the anisotropic ejecta pattern of the quark-gluon plasma, 
an asymmetric granular sheet was produced in an experiment using a granular jet with a rectangular cross-section \cite{5}.

\begin{figure}
\begin{center}
\includegraphics[width=3.35in]{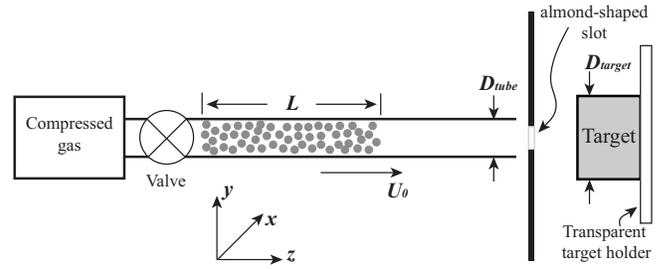}
\end{center}
\caption[experiment setup]{A schematic of the experiment setup. The
orientation of the coordinate system is also shown.} \label{Figure1}
\end{figure}

A crucial feature of RHIC experiment is that the degree of
anisotropy in the ejecta pattern varies with the impact parameter
between two ions: while a head-on collision with zero impact
parameter results in symmetric ejecta, a glancing collision with
impact parameter close to the diameter of an ion leads to a highly
anisotropic elliptic flow \cite{9,14,15}.  Are the anisotropic ejecta
patterns a universal feature of fluid impact? Can one mimic the
variation of anisotropic ejecta in a classical fluid experiment? To
answer these questions, the experimental challenge is to create
a fluid jet with a non-circular cross-section of tunable aspect
ratio to simulate the collision zone of two
partially overlapped relativistic ions \cite{9,14}. The experiment
is difficult, if not impossible, to perform with normal liquids,
where surface tension causes strong oscillations in the shape of a non-cylindrical jet and thus leads to uncontrollable collision
areas.  Here, we employ a granular material as a special fluid with
negligible surface tension \cite{16,17}. We perform a series of
experiments using non-cylindrical granular jets with
the aspect ratio of their cross-section ranging from 1 up to 6. We uncover two distinct regimes of jet impact. Moreover, we show that the anisotropy of ejected
granular sheet increases with the jet aspect ratio.  Our results provide
a new insight into the collective liquid-like behavior of granular
materials and demonstrate a universal feature of the fluid-impact
process. Beyond its academic interest as an analog to RHIC
experiments, the results indicate a new way to manipulate the
impact dynamics of granular jets, which have been widely used
in industrial processes such as abrasive machining, sand-blast
cleaning and polishing \cite{18,19}.


\section{II. Experiment}

Our experimental setup is similar to that used in previous studies
\cite{5,11}.  It consists of a glass launching tube filled with a
densely-packed granular material (Fig.~\ref{Figure1}).  The tube has
an inner diameter of $D_{tube} = 15$ mm, and is connected to a
high-pressure gas tank through a valve.  The gas accelerates a granular column of length $L = 40$ cm to a
speed $U_0 \sim 10$ m/s at the opening of the launching tube along the $z$ direction. During the impact, the speed of the granular jet varies: the jet slows down initially due to the impact and then accelerates as the length of the granular column becomes shorter. However, as shown previously, the dynamics of granular jet impact is independent of the impact speed due to the nearly zero surface-tension of granular matter \cite{5}. 

The solid target has a diameter $D_{target} = 1.8$ cm. We choose a large target such that the area of the target $\pi D_{target}^2/4$ is 4.4 times larger than the cross-section of granular jets. The target is centered on the jet axis and is 1.4 cm away from a changeable slot (Fig.~\ref{Figure1}). To modify the jet cross-section, a slot of prescribed shape and aspect ratio is inserted between the tube and the target. After passing through the slot, the jet acquires a cross-section with a shape approximately the same as the slot.  By
inserting different slots, we can control the shape of the jet's
cross-section.  A small gap of about 5 mm is left between the slot
and the tube to avoid jamming of the moving jet inside the tube. After passing the slot, the impact speed of granular column decreases to $U_0 \sim 6.0$ m/s. A
high-speed camera (Phantom v7) was used to image the impact dynamics
at 2000 frames per second from behind the target. To improve imaging, the slot is surrounded by a black screen so that only particles behind the slot could be viewed. Furthermore, we constructed a transparent target and target holder for some experiments so that the ejected particles can be seen without obstruction. 

Throughout our experiment, we use glass beads of diameters $90.5\pm15.5$
$\mu$m as our granular material (Mo-Sci Corporation, Class IV, Sieve Mesh Size -140+200). By tapping the launching tube, the beads are compacted close to the random-close packing density before each run.

\begin{figure}
\begin{center}
\includegraphics[width=2.6in]{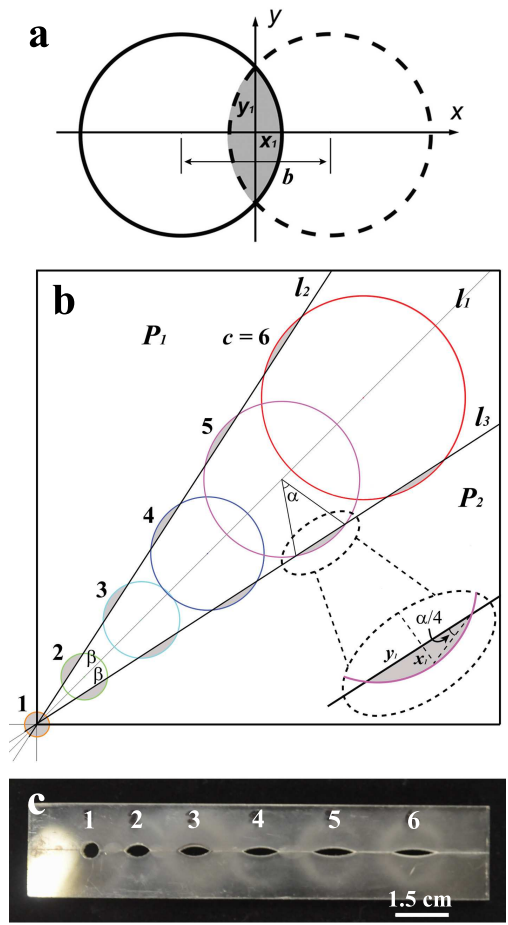}
\end{center}
\caption[Shape of granular jet's cross-section]{(color online). Shape of granular jet's cross-section. (a) Collision zone between two spherical ions (gray area). The left ion (solid line) moves in the z direction out of the plane of the paper. The right ion (dashed line) moves in the negative $z$ direction into the plane of the paper. The spherical ions are subjected to Lorentz contraction in the $z$ direction and become disks in the laboratory frame (see Section III C). $x_1$ and $y_1$ define the semi-axis of the collision zone. $b$ is the impact parameter.   (b) Schematic of the protocol for machining slots of different aspect ratios. The relation between
$\alpha$ and $c = y_1/x_1$ is shown in the lower right inset. (c) Slots of different aspect ratios. The dimension of each slot is given in Table~\ref{tab:table1}.} \label{Figure2}
\end{figure}

Since one of our goals is to mimic the collision zone of the RHIC experiments, the choice of the jets' cross-section shape is crucial. The azimuthal ($x-y$) projection of the collision zone between two
spherical ions consists of two circular segments with an
almond-like shape (Fig.~\ref{Figure2}a) \cite{9,14}.  The aspect
ratio of the zone, $c = y_1/x_1$, depends on the impact parameter,
$b$, and the size of the ions. Accordingly, we use a series of slots to produce granular jets with cross-sections similar to the collision zone of RHIC experiments. To manufacture slots of different aspect ratios, we adopt the following machining protocol.  First, holes of different sizes are drilled
along a straight line, $l_1$, on two identical 6.0 mm thick polycarbonate plates
(Fig.~\ref{Figure2}b), where the radius of the $i$-th hole, $R_i$, and its
center location along $l_1$, $D_i$, are determined by:
\begin{equation}
\label{R} {R_i=\left(\frac{\sigma}{\alpha_i-\sin\alpha_i}\right)^{1/2}},
\end{equation}
and
\begin{equation}
\label{D} {D_i=\frac{R_i\cos(\alpha_i/2)}{\sin\beta}}.
\end{equation}
Here, $\alpha_i = 4\arctan(1/c_i)$, $\sigma$ is the desired area of the
slot, and $\beta$ is an arbitrary angle that one chooses to cut the
plates in the next step. Specifically, we drill six holes on each plate with $c_i =$
1, 2, 3, 4, 5 and 6 respectively and choose $\beta = 12^\circ$. Next, we cut the first
plate along line $l_2$ and cut the second plate along line $l_3$. Each of these two lines forms an angle $\beta$ with respect to the central line $l_1$ (Fig.~\ref{Figure2}b). Joining the part $P_1$ with the part $P_2$ from these two cuts, we obtain six slots depicted as the gray areas in Fig.~\ref{Figure2}b. The final slots are shown in Fig.~\ref{Figure2}c and the dimension of the slots are indicated in Table~\ref{tab:table1}. Using these slots, we can generate granular jets with cross-sections similar to the collision zones of relativistic heavy ions at different impact parameters.  A further discussion of the analogy between our experiments and RHIC experiments can be found in Section III C.  

\begin{table}
\caption{\label{tab:table1} Dimension of the almond-shaped slots used in the experiments.}
\begin{ruledtabular}
\begin{tabular}{cccccccc}
 $c$ & $\sigma$ (mm$^2)$ & $R$ (mm) & $D$ (mm)& $x_1$ (mm) & $y_1$ (mm)\\
\hline
1& 16.6 & 2.30 & 0 & 2.30 & 2.30 \\ 

2& 16.6 & 4.31 & 12.43 & 1.72 &  3.45 \\ 

3& 16.6 & 7.12 & 27.41 &  1.43 & 4.27 \\ 

4& 16.6 & 10.54 & 44.72 & 1.24 & 4.96 \\ 

5& 16.6 & 14.45 & 64.15 & 1.11 & 5.56 \\ 

6& 16.6 & 18.79 & 85.50 & 1.02 & 6.09 \\ 

2\footnotemark[1]& 25.8 & -- & -- &  1.83 & 3.65 \\ 

\end{tabular}
\end{ruledtabular}
\footnotetext[1]{This large slot was made using a different method. Instead of the almond-like shape, the slot has an approximate rectangular shape with rounded corners. The slot had been used before in our previous study \cite{5}.}
\end{table}

To ensure that the granular jets passing through a slot are not perturbed too much in their motions and acquire a cross-section with the same shape as the slot, we performed two experiments. First, we removed the target and observed the jets moving directly toward the camera. After passing through the slot, the particles still move along the axis of the launching tube with only a slight downward bending due to the gravity. Therefore, the slots do not induce significant particle motion normal to the axis of the jets. Although the axial velocity of the particles may be modified after particles pass through the slot, the impact dynamics are not affected by the magnitude of impact speed \cite{5}. We performed another experiment to quantitatively measure the shape of the cross-section of a granular jet by detecting particle motion immediately in front of a transparent target. From this measurement, we confirm that the jet cross-section is quantitatively similar to the shape of the inserted slot (see Section III B).

\begin{figure}[t]
\begin{center}
\includegraphics[width=3.35in]{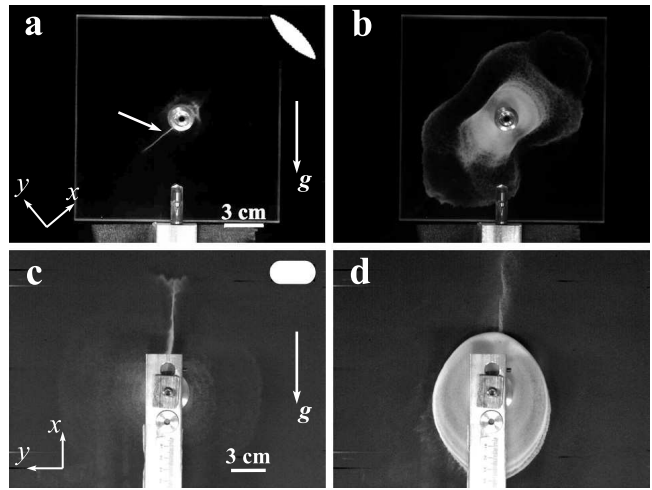}
\end{center}
\caption[Impact dynamics of granular jet]{Impact dynamics of
granular jets with $\sigma = 16.6$ mm$^2$ and $c = 4$ (a, b) and $\sigma = 25.8$ mm$^2$ and $c = 2$ (c, d). (a) and (c): First
transient regime; (b) and (d): Second steady-state regime. The focused stream in (a) is indicated by an arrow. In (c) and (d), the target holder partially blocks the view of granular ejecta. A downward moving stream can be seen from the side-view of the impact process. The cross-section and orientation of the slots are indicated at the upper left corner of (a) and (c). The linear size of real slots is 5 times smaller than those shown. The long axis of the slot is along the $y$ axis as indicated in (a) and (c).} \label{Figure3}
\end{figure}


\section{III. Results and Discussion}

Throughout our experiments, we maintain the area of the target 4.4 times larger than the cross-section of jets. Hence, the ejected granular sheet is always perpendicular to the incoming jet. This is consistent with a previous study, where granular cones with apex angle $< 90^\circ$ sets in only when $D_{target}/D_{jet} \le 1.8$ \cite{5}. In this paper, we focus on the shape of the ejected granular sheet in the $x-y$ plane.

\subsection{A. Dynamics of granular jet impact}

The dynamics of granular-jet impact shows two distinct regimes.
We illustrate these regimes with a small jet of $\sigma = 16.6$ mm$^2$, $c = 4$ and a large jet of $\sigma=25.8$ mm$^2$, $c=2$. When the jet first impacts the target, the energy of the spreading ejecta is highly focused. The jet deforms into one or two thick streams, which emerge along the short axis (the $x$ axis) of the jet's
cross-section (Fig.~\ref{Figure3}a, c). For jets with a small cross-section, the streams are weak and asymmetric (Fig.~\ref{Figure3}a and Supplementary Movie 1 \cite{20}). For jets with a larger cross-section, two strongly focused streams are observed, which emerge in opposite directions along $x$ (Fig.~\ref{Figure3}c and Supplementary Movie 2 \cite{20}). The target holder partially blocks the view of the ejecta in Fig.~\ref{Figure3}c, d. From a side-view of the impact, one can identify a downward moving stream symmetric to the upward shooting one shown in Fig.~\ref{Figure3}c. 

This first regime of granular impact is a transient state. As shown recently, during impact particles pile up in front of the target into a ``dead zone'', where particle mobility goes to zero \cite{11}.  The ejection of streams in this regime may be related to the formation of this ``dead zone'' during initial impact. For a jet with smaller cross-section ($\sigma = 16.6$ mm$^2$), this transit regime is shorter and less pronounced (Fig.~\ref{Figure3}a and Supplemental Movie 1 \cite{20}). 

In the second regime, after the emergence of the thin streams, an asymmetric granular sheet emerges (Fig.~\ref{Figure3}b and d). The major axis of the sheet is along the short axis of the impacting jet's cross-section.  The anisotropy of the sheet implies that particles that escape along the short axis of the collision zone have larger momentum. 

In this second regime, the granular jet reaches a steady state.  The ejecta is less focused, and the pattern persists over a longer time. Increasing $L$, the length of jet, increases the duration. Both the anisotropic shape and its relative orientation in Fig.~\ref{Figure3}d are reminiscent of the anisotropic pattern of strongly interacting Fermi gas atoms, where the interaction between individual atoms also leads to collective liquid-like behavior \cite{26}. In the rest of the paper, we will focus on the steady second regime of impact. 

\begin{figure}
\begin{center}
\includegraphics[width=3.35in]{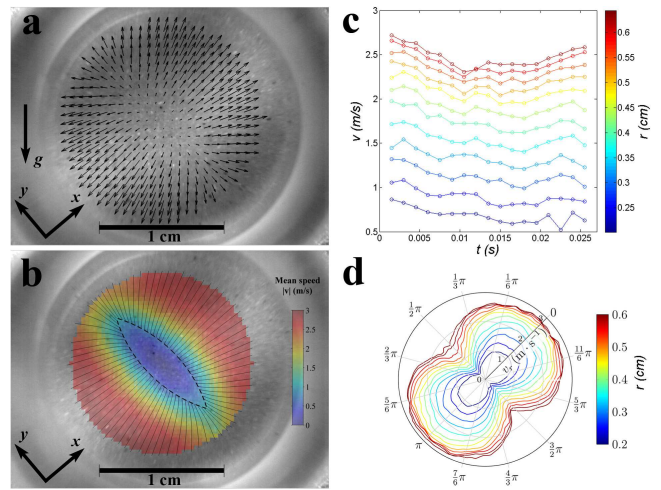}
\end{center}
\caption[Particle motion behind target]{(color online). Particle motion behind a transparent target with $\sigma = 16.6$ mm$^2$ and $c = 4$.  (a) Mean velocity field in the steady-state regime from the PIV measurement. (b) Mean radial speed in the azimuthal plane (in color). For comparison, the shape and orientation of a slot with $c=4$ (dashed line) is overlaid on the image.  The direction of gravity is indicated in (a). Streamlines from the velocity field in (a) are displayed. (see also Supplementary Movie 4 \cite{20}). (c) Radial velocity of particles at different radial locations, $r$, as a function of time. (d) Azimuthal distribution of the mean radial velocity of particles for circular rings at different $r$. $r$ is indicated by the color of the lines. The magnitude of velocity is indicated by the polar axis of the plot. A time average of 27.5 ms was used.} \label{Figure4}
\end{figure}

\subsection{B. Anisotropic ejecta}

The targets act to reflect the impact momentum of the impinging jet. The coordinates are defined in Fig.~\ref{Figure1}: the incoming granular jet moves in the $+z$ direction. The $x-y$ plane defines the azimuthal plane, and the $x$ and $y$ axes are along the short and long axes of the slot respectively. $\phi$ is the azimuthal angle in the $x-y$ plane and $\theta$ is the polar angle out of the $x-y$ plane. For simplicity, we first neglect the energy loss due to the inelasticity of particle collision and assume the particles are spheres. After impact, the reflected particles move in the $-z$ direction, collide with other incoming particles in the jet, acquire a velocity component in the azimuthal direction normal to the $z$ axis, and eventually escape the impact zone. The frequency and duration of particle collisions depend on the concentration of particles in the jet. For very concentrated granular jets, persistent contact between particles may occur. Structures such as ``dead zones'' emerge during impact, which can be understood in terms of incompressible granular fluids \cite{11}. Accordingly, two limiting cases should be considered: 

(1) In the ideal gas limit, the particles in the jet are dilute so that the chance of a particle colliding a second time with another particle is rare (Supplementary Movie 3 \cite{20}). Most reflected particles collide only once and then move ballistically out of the impact zone. Since the relative position between a reflected particle and an incoming one is random, the collision-induced azimuthal velocity component is uniformly distributed. In other words, the scattering cross-section is independent of $\phi$. A geometric argument shows that the distribution of particles in the polar angle direction is proportional to $\sin(\theta)$ \cite{21}. Since the impact process only involves uncorrelated two-particle collisions, it does not depend on the shape of collision zone; a symmetric ejecta will form regardless of the shape of the cross-section of the jet. Adding inelasticity and non-spherical shapes of realistic granular particles to the problem does not affect the above conclusion. Symmetric ejecta in low-density jets were shown in previous work \cite{5}.

(2) As the particle density increases, a particle experiences many more collisions. The mean free path becomes smaller than the particle size when the concentration is high \cite{5}. In this limit, a large number of collisions creates a highly-interacting region \cite{22,23} so that a hydrodynamic pressure builds up in the center of the impact zone. The pressure gradient is largest along the $x$ axis, where the distance between the center and the boundary is shortest. This pressure gradient induces the highest particle momentum along the short axis and leads to the observed anisotropic ejecta. The inelastic nature of the collisions may also lead to particle clusters in an isolated system \cite{24,25}. Such a cluster instability is not observed in our experiment, where the system has an open boundary with a rapid energy influx.

We directly probe the formation of the anisotropic ejecta in the second steady-state regime by investigating the detailed particle dynamics behind a transparent target (Fig.~\ref{Figure4} and Supplementary Movie 4 \cite{20}). The zoom-in video allows us to track the motion of particles using Particle Imaging Velocimetry (PIV). Figure ~\ref{Figure4} shows the steady-state velocity field for a granular jet passing through a slot of $c=4$. As expected, the constant velocity contours have an almond-like shape quantitatively similar to the shape of the slot (Fig.~\ref{Figure4}b). The aspect ratio of the constant velocity profiles is $3.6\pm0.4$ near the impact point, which is very close to the aspect ratio of the slot, $c=4$. The radial velocity of particles varies slightly in the steady state (Fig.~\ref{Figure4}c). Particle velocity at $r=6.0$ mm away from the center of the impact zone decreases first upon impact and increases toward the end of the steady state. The overall change is about $15\%$. Fig.~\ref{Figure4}d shows the radial velocity distribution of particles at different radial distances from the impact-zone center. Note that the velocity is largest along the short axis of the slot, consistent with the large-scale measurement shown in Fig.~\ref{Figure3}b. More important, the measurement shows that the motion of particles quickly converge to a saturated anisotropic distribution within 4 ms after impact . The velocity approaches a steady profile between 5 $\sim$ 6 mm (50 to 60 particles diameters) from the center of the impact zone (Fig.~\ref{Figure4}d). This central region where particle velocities increases from almost zero to 3 m/s is an almond-shaped ``dead zone'', similar to the symmetric ``dead zone'' found in the impact of cylindrical granular jets \cite{11}.

\subsection{C. Analogy to RHIC experiment}

The asymmetric ejecta pattern has been used as critical evidence for the liquid-like behavior of the quark-gluon plasma in the RHIC experiments \cite{8,9,14}. The analogy between those experiments and the granular-jet experiments can be seen from the geometry of the impact process. As explained in the previous section, the target in our experiment acts as a mirror to reflect particle momentum. Hence, the shape of the collision zone where the reflected particles collide with the incoming particles is the same as the cross-section of the granular jet. In the RHIC experiments, due to relativistic effects, spherical ions become disks of thickness of $R/\gamma$ in the lab frame, where $R$ is the radius of heavy ions in the rest frame, $\gamma = 1/\sqrt{1-v^2/c^2}$ is the Lorentz factor, $v$ is the velocity of the impacting ions and $c$ is the speed of light in vacuum \cite{14}. The thickness of the disk $R/\gamma$ plays the role of the length of the granular jets in our experiments. The collision zone -- the partially overlapped area of two disks -- has the same shape as in granular jet experiments (Fig.~\ref{Figure2}a). In a RHIC collision, there exists a time delay between the impact and the formation of ejecta \cite{8,14}. Hence, the parts of heavy ions outside the overlapped area, which pass by the collision zone after the impact with a speed near the speed of light, do not affect the formation of ejecta.

Clearly, the nature of interactions in the RHIC and in the granular-jet experiments are completely different. However, the fact that the RHIC results can be successfully interpreted by a coarse-grained, long-wavelength hydrodynamic theory indicates that the impact dynamics are largely independent of detailed interactions \cite{14}. Hence, it is not surprising that our experiments show similar ejecta patterns as the RHIC experiments. A quantitative comparison of the shape of ejecta in the two experiments will be shown in Section III E.

\begin{figure}
\begin{center}
\includegraphics[width=3.35in]{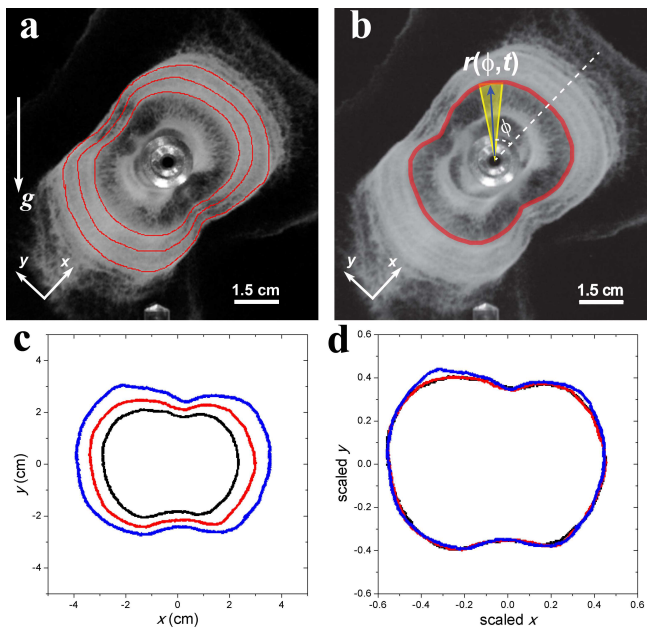}
\end{center}
\caption[Time dependence of Anisotropy]{(color online). Quantitative measurement of ejecta dynamics. (a) Three concentric material rings (red curves) in the ejecta sheet.  The image is from a granular jet with $\sigma = 16.6$ mm$^2$ and $c = 4$. (b) A schematic showing the quantitative measurement of $r(\phi)$ and $P(\phi)$. The thick red line indicates the shape of the material ring at the end of the second regime. The blue arrow indicates the location of the ring, $r(\phi)$. The yellow wedge indicates the differential area around $\phi$, $\mathrm{d} A$, which is proportional to $P(\phi)$ (see the main text). (c) The material rings extracted from (a). (d) Scaling of the three rings by their characteristic lengths.}
\label{Figure5}
\end{figure}

\begin{figure}
\begin{center}
\includegraphics[width=3in]{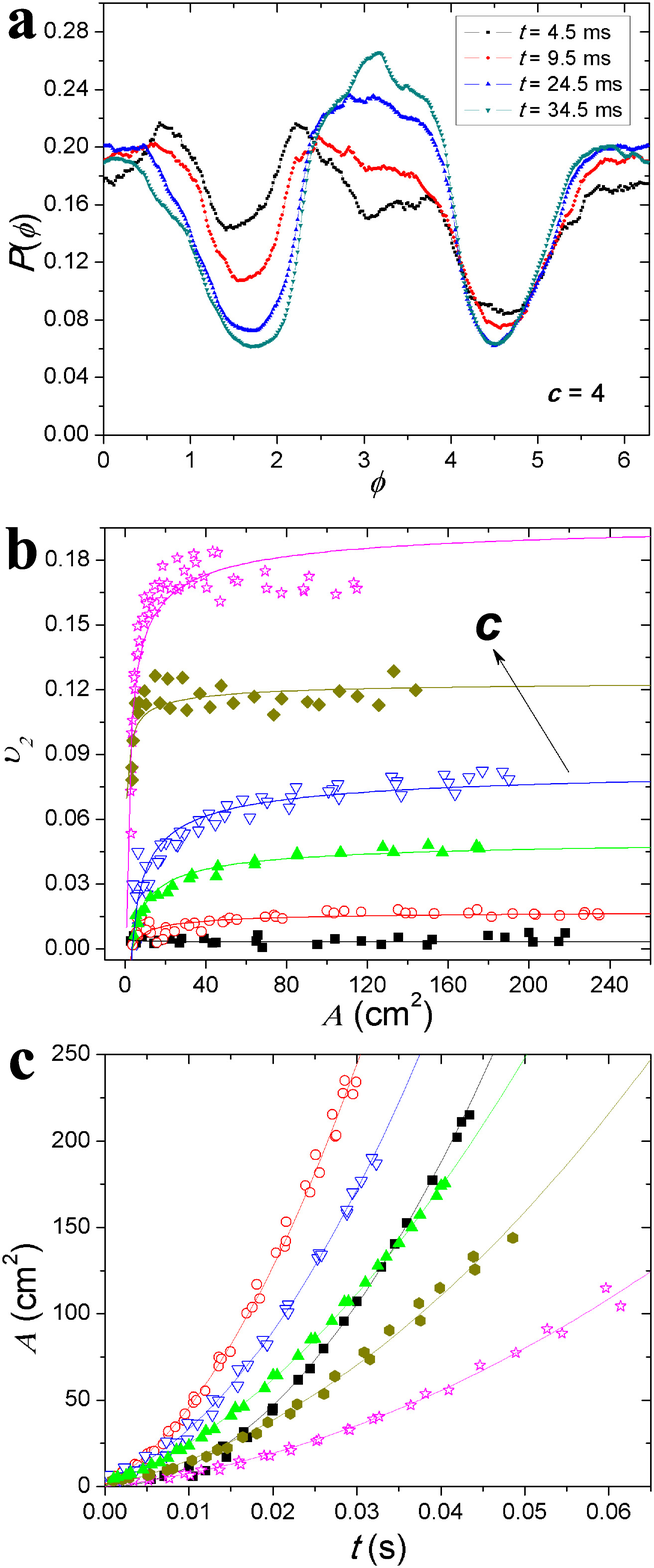}
\end{center}
\caption[Time dependence of Anisotropy]{(color online). Dynamics of anisotropic ejecta. (a) Azimuthal momentum distribution $P(\phi)$ at different $t$ for the granular jet of $\sigma = 16.6$ mm$^2$ and $c = 4$. (b) Second Fourier coefficient of $P(\phi)$, $\upsilon_2$, as a function of $t$ for granular jets of different $c$. Since the area enclosed inside a material ring, $A$, increases monotonically with $t$ [see (c)], we use $A$ as the time axis. The aspect ratio, $c$, of experimental curves increases from 1 to 6 from bottom to top. The solid lines are square-root fits (Eq.~\ref{equ5}). (c) Enclosed area inside a material ring, $A$, as a function of $t$. Different curves are for the jets of different $c$. The symbols are the same as those used in (b). Solid lines are second-order polynomial fits (Eq.~\ref{equ2}).}
\label{Figure6}
\end{figure}

\subsection{D. Quantitative analysis of anisotropic ejecta: $\upsilon_2$}

Due to a variation in particle density in the incoming jet, a series of concentric material rings can be identified on the ejecta sheet in the second regime of impact process. Such features indicate different groups of particles that impact on the target at different times and can be employed to track the dynamics of the ejecta (Fig.~\ref{Figure5}a and Supplementary Movie 1 \cite{20}). As the most obvious example, the trailing edge of the ejecta sheet arises from the impact of the last group of particles on the target (Fig.~\ref{Figure5}b). We investigate the dynamics of ejecta by tracking the radial position of these material rings, $r(\phi,t)$ (Fig.~\ref{Figure5}b). Here, $t$ is defined as the time elapsed after the impact of the group of particles that creates the specific ring under consideration. 

Different rings have different radial velocities -- a ring created at later times tends to have larger radial velocities due to the increasing pressure gradient across the jet and hence it will sweep out all the rings in front of it during its outward motion (Supplementary Movie 1 and 2 \cite{20}).  However, at a fixed time, rings at different radial locations can be collapsed onto a master curve if scaled by the characteristic size of the rings (Fig.~\ref{Figure5}c,d). Hence, any of these concentric rings can be used for studying the shape of ejecta when a proper normalization is implemented. Practically, it is more convenient to choose the trailing edge of the ejecta (Fig.~\ref{Figure5}b), since it has the maximum velocity and the sharpest contrast to the background throughout the impact process (Supplementary Movie 1 \cite{20}).

The physical meaning of $r(\phi)$ at a given time $t$ can be understood as follows. After the velocity of particles converges to a steady profile which happens within a small region surrounding the ``dead zone'' (Fig.~\ref{Figure4}), particles in the ejecta simply move ballistically outward from the center of the impact zone. Therefore, at large distances, the radial location of a ring, $r(\phi)$, is related to the radial particle velocity $v_p(\phi)$ simply by $r(\phi) \approx v_p(\phi)t$. Hence, $r(\phi)$ reveals the particles' velocity distribution $v_p(\phi)$. More importantly, if we assume that the particle flux $q$ along $\phi$ is proportional to the particle velocity along the same direction $q(\phi) \sim v_p(\phi)$, then the total particle momentum in the $\phi$ direction, $P(\phi) = \sum{mv_p(\phi)} \sim q(\phi)v_p(\phi) \sim v_p(\phi)^2 \sim r(\phi)^2$, where $m$ is the mass of individual particles and the summation is over all the particles moving in the $\phi$ direction in a specific material ring.  In other words, the momentum distribution of the ejecta, $P(\phi)$, is proportional to the differential area of the plane enclosed by the material ring under consideration, $\mathrm{d} A = \frac{1}{2}r(\phi)^2 \mathrm{d}\phi$. The area can be directly measured from the impact movie (Fig.~\ref{Figure5}b). The proportionality constant between $P(\phi) \mathrm{d}\phi$ and $\mathrm{d} A$ can be determined by requiring $\int_0^{2 \pi} P(\phi) \mathrm{d} \phi=1$. This normalization condition leads to $P(\phi) = \frac{1}{2}r(\phi,t)^2/A$, which eliminates the dependence of $P(\phi)$ on the total area enclosed by the material ring $A = \int_0^{2 \pi} \mathrm{d} A = \frac{1}{2} \int_0^{2 \pi} r(\phi)^2 \mathrm{d} \phi$. The azimuthal momentum distribution of ejecta, $P(\phi)$, is one of the most important parameters charactering impact process and has been measured in the RHIC experiments \cite{8,14,27}.

The dynamics of the momentum distribution $P(\phi)$ can thus be obtained from impact movies (Fig.~\ref{Figure6}a). To quantify the degree of ejecta's anisotropy, we follow the same procedure previously used in the RHIC experiments and in the study of granular jets \cite{5,8}: the second coefficient of the Fourier expansion of $P(\phi)$,  $\upsilon_2 = \frac{1}{2\pi} \left|  \int_0^{2 \pi} P(\phi)e^{-i2\phi} \mathrm{d} \phi \right|$ is calculated. A large $\upsilon_2$ indicates a strong anisotropic structure.   

The dynamics of $\upsilon_2$ for granular jets of different aspect ratios is shown in Fig.~\ref{Figure6}b. For all the aspect ratios, the anisotropy of the granular ejecta increases with time. The sheet is more isotropic when it first emerges as indicated by the small $\upsilon_2$ close to 0. The degree of the anisotropy increases monotonically and quickly approaches a plateau. Since the total area enclosed by the material ring, $A$, increases monotonically with $t$ as shown in Fig.~\ref{Figure6}c, we use $A$ to indicate the increase of time in Fig.~\ref{Figure6}b. 

The increase of ejecta's anisotropy can be understood by considering a more accurate ballistic model, in which the radial location of a particle ring is  given by: 
\begin{equation}
\label{equ1} {r(\phi,t) = r_0 + v_p(\phi)t}.
\end{equation} 
Here, $r_0$ represents the size of the region over which particles adapt to the final ejecting velocity (Fig.~\ref{Figure4}d). $r_0$ can be considered as the size of ``dead zones'' \cite{28}. Since $r_0$ is only about a few millimeters, $r(\phi,t) \approx v_p(\phi)t$ when $t$ is large, in consistence with our interpretation of the relation between $r(\phi)$ and $P(\phi)$ shown above. The total area enclosed by the ring is then:
\begin{eqnarray}
A(t) & = & \frac{1}{2}\int_0^{2 \pi} r(\phi)^2 \mathrm{d} \phi \nonumber \\
& = & \pi r_0^2 + r_0\langle v_p(\phi)\rangle t + \frac{1}{2}\langle v_p(\phi)^2\rangle t^2,
\label{equ2}
\end{eqnarray}  
where for any function $f(\phi)$, $\langle f(\phi) \rangle = \int_0^{2 \pi} f(\phi) \mathrm{d} \phi$. 
We fit the experimental data $A(t)$ with Eq.~\ref{equ2} (Fig.~\ref{Figure6}c). The fitting leads to $r_0 = 4.2 \pm 3.1$ mm, consistent with the direct measurement behind the transparent target (Fig.~\ref{Figure4}d). The first coefficient, $\langle v_p(\phi)\rangle$, and the second coefficient, $\langle v_p(\phi)^2\rangle$, are fitting parameters depending on the absolute velocity of the ring, which varies between different measurements. However, the ratio between the two, $\langle v_p(\phi)^2\rangle / \langle v_p(\phi)\rangle$, is always larger than 1, ranging from 4 up to 65 cm/s. Therefore, the second order term quickly dominates at large $t$.   

The normalized momentum distribution is $P(\phi,t) = \frac{1}{2}r(\phi,t)^2/A(t)$. Its second Fourier coefficient is given by:
\begin{eqnarray}
\upsilon_2 & = & \frac{1}{2 \pi A(t)}\left| \int_0^{2 \pi} \frac{1}{2} r(\phi)^2 e^{-i2\phi} \mathrm{d} \phi \right|  \\
& = & \frac{1}{2 \pi A(t)}\left[ r_0\left| \langle v_p(\phi)e^{-i2\phi}\rangle \right| t + \frac{1}{2}\left| \langle v_p(\phi)^2 e^{-i2\phi} \rangle \right| t^2 \right]. \nonumber
\label{equ3}
\end{eqnarray} 
Inserting Eq.~\ref{equ2} into Eq.~\ref{equ3}, at large $t$ limit, it becomes:
\begin{widetext}
\begin{eqnarray} 
\upsilon_2 & = & \frac{1}{2\pi} \left[ \frac{\left| \langle v_p(\phi)^2 e^{-i2\phi}\rangle \right|}{\langle v_p(\phi)^2\rangle} - \frac{2r_0}{\langle v_p(\phi)^2\rangle} \left(\frac{\langle v_p(\phi) \rangle \left| \langle v_p(\phi)^2 e^{-i2\phi}\rangle \right|}{\langle v_p(\phi)^2\rangle} - \left| \langle v_p(\phi)e^{-2i\phi}\rangle \right|\right)\frac{1}{t}\right] \nonumber \\
& = & \frac{1}{2\pi} \left[ \frac{\left| \langle v_p(\phi)^2 e^{-i2\phi}\rangle \right|}{\langle v_p(\phi)^2\rangle} - \frac{2r_0}{\langle v_p(\phi)^2\rangle} \left(\frac{\langle v_p(\phi) \rangle \left| \langle v_p(\phi)^2 e^{-i2\phi}\rangle \right|}{\langle v_p(\phi)^2\rangle} - \left| \langle v_p(\phi)e^{-2i\phi}\rangle \right|\right)\sqrt{\frac{\langle v_p(\phi)^2 \rangle}{2A}}\right].
 \label{equ4}
\end{eqnarray}
\end{widetext}  
In the last step, we use $t = \sqrt{2A(t)/\langle v_p(\phi)^2 \rangle}$ in the large $t$ limit. Hence, we find:
\begin{equation}
\label{equ5} {\upsilon_2 = a - bA^{-\frac{1}{2}}},
\end{equation}
where $a$ and $b$ are fitting parameters depending on the velocity distribution of ejecta and its second moment. The relation fits well with our experimental data as shown in Fig.~\ref{Figure6}b.  

\begin{figure}
\begin{center}
\includegraphics[width=3.35in]{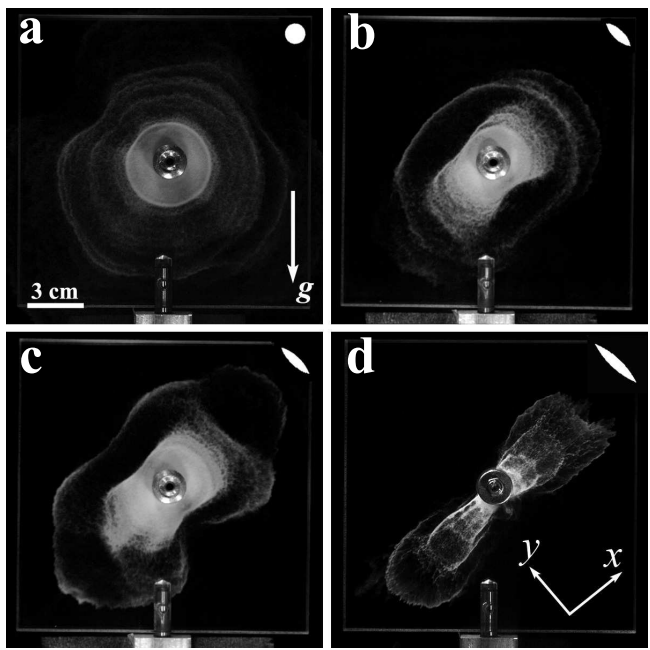}
\end{center}
\caption[Impact of granular jets with different aspect
ratios]{Impact of granular jets with cross-section area $\sigma = 16.6$ mm$^2$ and aspect ratio: (a) $c = 1$, (b) $c = 3$, (c) $c = 4$, and (d) $c = 6$. The cross-section and orientation of the slots are indicated at the upper left corner of each plot. The linear size of real slots is 2.5 times smaller than those shown in the plot.}
\label{Figure7}
\end{figure}

\begin{figure}
\begin{center}
\includegraphics[width=3.35in]{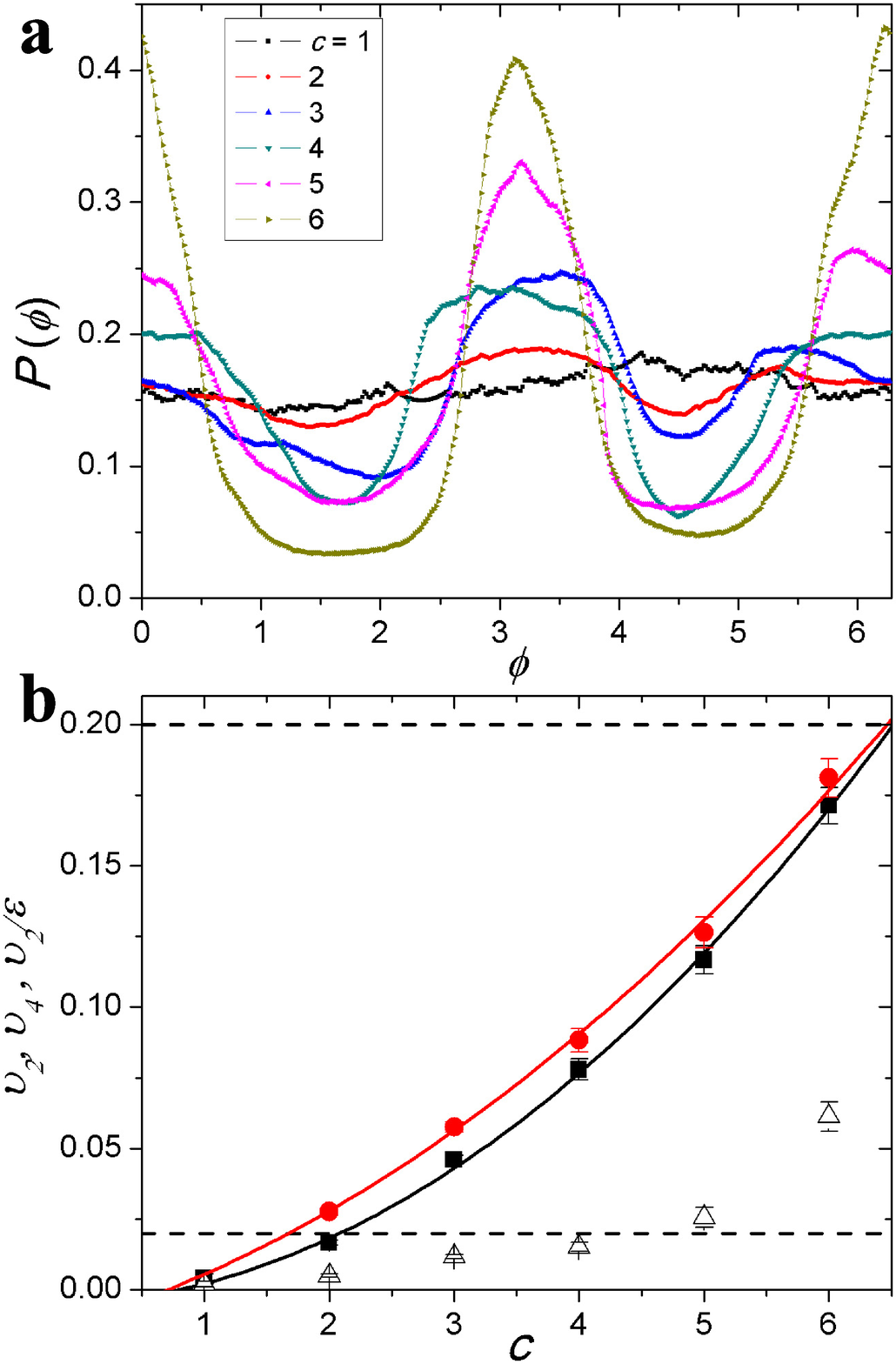}
\end{center}
\caption[Shape of granular sheets for jets of different
cross-sections]{(color online). Azimuthal momentum distribution and the anisotropy of ejecta for granular jets of cross-section area $\sigma = 16.6$ mm$^2$ and different aspect ratios.  (a) Azimuthal momentum distribution, $P(\phi)$, for granular jets of different $c$.  The distribution is taken from the large $t$ limit where the ejecta has achieved a steady shape. (b) Anisotropy $\upsilon_2$ (black squares) and higher harmonic anisotropy,  $\upsilon_4$ (triangle) as a function of $c$. $\upsilon_2/\varepsilon$ is also shown (red disks). The data are averaged over three to four experimental trials. Solid lines are polynomial fits to guide eyes. The horizontal dashed lines indicate the upper and lower limits of $\upsilon_2/\varepsilon$ observed in the RHIC experiments \cite{27}.} \label{Figure8}
\end{figure}

\subsection{E. Aspect ratio dependence of ejecta}

Finally, we investigate the dependence of the ejecta on the aspect ratio of the granular jets. Since the anisotropy approaches an asymptotic value at large $t$ (Fig.~\ref{Figure6}b), we examine the shape of the ejecta in this large $t$ limit for jets of different $c$. The asymptotic anisotropy strongly depends on the aspect ratio of jet's cross-section.  As expected, the jet with $c = 1$ shows a symmetric
ejecta pattern (Fig.~\ref{Figure7}a).  With $c = 1$, one cannot
distinguish between particle-like and liquid-like behaviors. At $c =
3$, however, the pattern is less symmetric (Fig.~\ref{Figure7}b). It
turns into a figure-eight shape with a narrow waist when $c = 4$ (Fig.~\ref{Figure7}c
and Supplemental Movie 1 \cite{20}). At $c = 6$,
the shape of the ejecta is elongated into a thin strip
(Fig.~\ref{Figure7}d). 

Quantitatively, we measure the azimuthal momentum distribution of ejecta $P(\phi)$ for
different $c$ (Fig.~\ref{Figure8}a). The slightly asymmetric shape between the two peaks at $\phi = 0$ and $\phi = \pi$ are due to the influence of gravity. The corresponding second Fourier coefficients of $P(\phi)$ are plotted in Fig.~\ref{Figure8}b.
The data clearly show the increase of the degree of anisotropy of
ejecta with the aspect ratio. From $c = 2$ to $c =
6$, $\upsilon_2$ increase monotonically from 0 to 0.18.     

We compare our results with the RHIC experiments quantitatively. In our experiments, $\upsilon_2$ varies between 0.02 and 0.18. This covers nearly the same range as the RHIC experiments which were between 0.028 and 0.10 at different transverse momenta \cite{29}. The RHIC experiments reported their results using a normalized quantity $\upsilon_2/\varepsilon$, where $\varepsilon = (c^2-1)/(c^2+1)$ is the geometric factor of the cross-section. We perform the same analysis for our experiments (Fig.~\ref{Figure8}b). Our $\upsilon_2/\varepsilon$ covers a range between 0.03 and 0.19, also overlapping well with the range between 0.02 and 0.20 of the RHIC results \cite{27}. In contrast with our previous study, where we chose the focused streams in the transient state as the analog to the RHIC results \cite{5}, here we analyze the anisotropy of the granular sheet in the second, steady-state, regime. In this regime, $\upsilon_2/\varepsilon$ is smaller. 

We also measure the higher harmonic, $\upsilon_4$ (Fig.~\ref{Figure8}b).  We find the ratio $\upsilon_2/\upsilon_4$ varies from 1.9 to 5.1 in our experiments. Similarly, in the RHIC experiments, $\upsilon_2/\upsilon_4$ varies from 2.8 to 6.0 \cite{29}. It is difficult to directly measure the impact parameter or the aspect ratio of the collision zone in RHIC experiments \cite{30}. Instead, the so-called centrality percentile is measured, which is a nonlinear function of the impact parameter. Although the general trend of RHIC results are consistent with our results, the functional forms of $\upsilon_2$ and $\upsilon_4$ versus $c$ appear to be significantly different \cite{29}.


\section{IV. Conclusions}

We have illustrated the liquid-like behavior of granular materials
by preforming granular-impact experiments with non-cylindrical jets. Using high-speed photography, we found much richer dynamics than previously reported \cite{5,6,7,10,11,12,13}. The collective hydrodynamics of granular jets is vividly demonstrated by the anisotropic shape of ejected sheets. Inspired by the RHIC experiments, we showed that the degree of anisotropy can be controlled by using jets of different aspect ratio. The simple ballistic motion of particles can be used to interpret the evolution of ejecta in the large-time limit. We find a quantitative similarity between our results and those from the RHIC experiments. The difference in the detailed shape of $\upsilon_2(c)$ and $\upsilon_4(c)$ may due to different geometries used in the two experiments. Rather than the collision of two jets, the RHIC experiments are better modeled by the collision of two spheres. Further experiments on the higher harmonics of azimuthal particle momentum distributions could be conducted in the future to reveal the transport coefficients of granular flows.   

Our experimental findings pose new questions for future studies.  How do granular particles equilibrate into a liquid state through collisions? Does the equipartition theorem apply in the granular impact process? Do different degrees of freedom, such as the translational and rotational motions of particles, have the same relaxation time?  What is the relation between the initial transient state of granular impact and granular-shock dynamics in loosely compacted granular media \cite{31,32,33,34,35}?  Current simulation techniques should be useful for investigating such highly non-equilibrium time-dependent processes \cite{10,11,12,13}? 

\section{Acknowledgements}

We are grateful to Ling-Nan Zou, Herv\'e Turlier and Daniel Citron,
R. Bellwied and S. Gavin for helpful advice.  This work was
supported by the NSF MRSEC DMR-0820054.  L.G. was supported by the
NSF Materials World Network DMR-0807012 and the AXA Research Fund.


\begin{thebibliography}{99}
\bibitem{28} For the sake of simplicity, we have chosen a circular material ring for $t=0$ instead of a general-shaped ring (i.e. $r_0$ as a function of $\phi$). Although a non-uniform $r_0(\phi)$ yields minor modifications to the formulas presented here, the general relation between $\upsilon_2$ and $A$ (Eq.~\ref{equ5}) is unchanged.
\end{thebibliography}
\end{document}